\documentclass[superscriptaddress,prl,twocolumn,nofootinbib,showpacs,preprintnumbers,floatfix]{revtex4}

\usepackage{amsfonts}
\usepackage{mathrsfs}
\usepackage{amssymb}
\usepackage{amsmath}
\usepackage{graphicx,epsfig}

\def\bwt{\begin{widetext}}

\def\ewt{\end{widetext}}

\def\be{\begin{equation}}

\def\ee{\end{equation}}

\def\bea{\begin{eqnarray}}

\def\eea{\end{eqnarray}}

\def\bean{\begin{eqnarray*}}

\def\eean{\end{eqnarray*}}

\def\bary{\begin{array}}

\def\eary{\end{array}}

\def\bit{\begin{itemize}}

\def\eit{\end{itemize}}

\def\su5u1{SU(5) \times U(1)}

\def\fsu5u1{SU(5) \times U(1)'}

\def\so10{SO(10)}

\def\sq20{SO(10) \times SO(10)}

\usepackage{amssymb}

\begin{document}

\setlength{\parskip}{0cm}

\title{The Superluminal Neutrinos from Deformed Lorentz Invariance}

\author{Yunjie Huo}

\affiliation{ State Key Laboratory of Theoretical Physics,
      Institute of Theoretical Physics, Chinese Academy of Sciences,
Beijing 100190, P. R. China }

\author{Tianjun Li}

\affiliation{ State Key Laboratory of Theoretical Physics,
      Institute of Theoretical Physics, Chinese Academy of Sciences,
Beijing 100190, P. R. China }

\affiliation{George P. and Cynthia W. Mitchell Institute for Fundamental Physics and Astronomy, Texas A$\&$M University, College Station, TX
77843, USA }

\author{Yi Liao}

\affiliation{ Center for High Energy Physics, Peking University, Beijing
100871, P. R. China }

\affiliation{School of Physics, Nankai University, Tianjin 300071, P. R. China} 

\author{Dimitri V. Nanopoulos}

\affiliation{George P. and Cynthia W. Mitchell Institute for Fundamental Physics and Astronomy,
 Texas A$\&$M University, College Station, TX 77843, USA }

\affiliation{Astroparticle Physics Group,
Houston Advanced Research Center (HARC),
Mitchell Campus, Woodlands, TX 77381, USA}

\affiliation{Academy of Athens, Division of Natural Sciences,
 28 Panepistimiou Avenue, Athens 10679, Greece }

\author{Yonghui Qi}

\affiliation{ State Key Laboratory of Theoretical Physics,
      Institute of Theoretical Physics, Chinese Academy of Sciences,
Beijing 100190, P. R. China }

\author{Fei Wang}

\affiliation{Department of Physics, Zhengzhou University, Zhengzhou, Henan, P. R. China }



\begin{abstract}

We study two superluminal neutrino scenarios 
where $\delta v\equiv (v-c)/c$ is a constant.
To be consistent with the OPERA, Borexino, and ICARUS experiments 
and with the SN1987a observations, we assume 
that $\delta v_{\nu}$ on the Earth is about three order larger 
than that on the interstellar scale. To explain the theoretical 
challenges from the Bremsstrahlung effects and pion decays, 
we consider the deformed Lorentz invariance, and show that 
the superluminal neutrino dispersion relations can be realized 
properly while the modifications to the dispersion relations of 
the other Standard Model particles can be negligible. In addition, 
we propose the deformed energy and momentum conservation laws for 
a generic physical process. In Scenario I the momentum conservation 
law is preserved while the energy conservation law is deformed. 
In Scenario II the energy conservation law is preserved while
the momentum conservation law is deformed. We present the energy 
and momentum conservation laws in terms of neutrino momentum in 
Scenario I and in terms of neutrino energy in Scenario II. 
In such formats, the energy and momentum conservation
laws are exactly the same as those in the traditional quantum 
field theory with Lorentz symmetry. Thus, all the above theoretical 
challenges can be automatically solved. We show explicitly that
the Bremsstrahlung processes are forbidden and there is no problem
for pion decays.

\end{abstract}


\preprint{ACT-20-11, MIFPA-11-52}

\maketitle


{\bf Introduction~--} Recently, the OPERA,  Borexino, and ICARUS
 experiments have measured
the muon neutrino ($\nu_\mu$) velocities with high accuracy through the 
measurement of the flight time and the distance (730 km) between the 
source of the CNGS neutrino beam at CERN (CERN Neutrino beam 
to Gran Sasso) and their detectors at the underground 
Gran Sasso Laboratory (LNGS)~\cite{Adam:2011zb, AlvarezSanchez:2012wg, Antonello:2012be}. 
At the OPERA experiment~\cite{Adam:2011zb}, the
arrival time of CNGS muon neutrinos with respect to the 
one for a particle moving at the speed of light in vacuum is 
\begin{eqnarray}
\delta t = \left( 6.5 \pm 7.4~ {\rm (stat.)}^{+8.3}_{-8.0} {\rm (sys.)} \right) {\rm ns}~.~\,
\end{eqnarray}
Thus, we obtain a relative difference of the muon neutrino velocity with 
respect to the speed of light
\begin{eqnarray}
\delta v_\nu &\equiv&  {{v_{\nu} -c}\over c} 
\nonumber \\
&=&
\left( 2.7 \pm 3.1 ~{\rm (stat.)}^{+3.4}_{-3.3} ~{\rm (sys.)} \right) \times 10^{-6}~,
\label{eq:opr}
\end{eqnarray}
The dependence of $\delta t$ on the neutrino energy was investigated as well
yielding a null effect.
At the Borexino experiment~\cite{AlvarezSanchez:2012wg}, we have 
\begin{eqnarray}
\delta t = \left( 0.8 \pm 0.7~ {\rm (stat.)} \pm 2.9 {\rm (sys.)} \right) {\rm ns}~,~\,
\end{eqnarray}
which is well consistent with zero.
At 90\% Confidence Level (C.L.), we get
\begin{eqnarray}
\delta v_\nu &<& 2.1\times 10^{-6} ~.~\,
\end{eqnarray}
Morever, at the ICARUS experiment~\cite{Antonello:2012be}, we have
\begin{eqnarray}
\delta t = \left( 0.18 \pm 0.69~ {\rm (stat.)} \pm 2.17 {\rm (sys.)} \right) {\rm ns}~,~\,
\end{eqnarray}
which is also well consistent with zero. At 95\% C.L., we get
\begin{eqnarray}
\delta v_\nu &<& 1.6\times 10^{-6} ~.~\,
\end{eqnarray}
To be concrete, we would like to summarize these experimental results:
(1) all the central values for $\delta t$ are non-zero and 
positive~\cite{Adam:2011zb, AlvarezSanchez:2012wg, Antonello:2012be}.;
(2) $\delta v_\nu $ around $10^{-6}$ is still consistent with 
all the experiments~\cite{Adam:2011zb, AlvarezSanchez:2012wg, Antonello:2012be}.;
(3) there is no energy denpendence for muon neutrino velocity~\cite{Adam:2011zb}.
Therefore, in this Letter we would like to study the possible superluminal 
 neutrinos with $\delta v_\nu $ around $10^{-6}$ which have no energy dependence.
From the theoretical point of view, many groups 
have already studied the possibility of superluminal 
neutrinos~\cite{Ellis:2008fc, Cacciapaglia:2011ax,
AmelinoCamelia:2011dx, Giudice:2011mm, Dvali:2011mn, Mann:2011rd, Drago:2011ua,
Li:2011ue, Pfeifer:2011ve, Lingli:2011yn, 
Alexandre:2011bu, Cohen:2011hx, GonzalezMestres:2011jc, Matone:2011jd, 
Ciuffoli:2011ji, Bi:2011nd, Wang:2011sz, Wang:2011sz-B, Cowsik:2011wv, Li:2011zm, 
AmelinoCamelia:2011bz, Moffat:2011ue, Faraggi:2011en, Konoplya:2011hf, Li:2011rt,
Zhao:2011sb, Chang:2011td, Almeida:2011ca, Matone:2011fn, Alles:2011wq, 
Evslin:2011vq, Lingli:2011kh, Mohanty:2011rm, Li:2011ad, Ling:2011re}.

The major challenges to the superluminal neutrino results are the following: 
(1) Bremsstrahlung
effects~\cite{Cohen:2011hx}. The superluminal muon neutrinos with $\delta v_{\nu}$ given in 
Eq.~(\ref{eq:opr}) would lose energy rapidly via Cherenkov-like processes
on their ways from CERN to LNGS, and the most important process is
 $\nu_{\mu} \to \nu_{\mu} + e^+ + e^-$. Thus, the experiment can not observe
the muon neutrinos with energy in excess of 12.5 GeV~\cite{Cohen:2011hx}; 
Considering such effects, the ICARUS experiment has set a tigh limit
$\delta v_\nu <1.25\times 10^{-8} $ at 90\% C.L.~\cite{ICARUS:2011aa}.
(2) Pion decays~\cite{GonzalezMestres:2011jc, Bi:2011nd, Cowsik:2011wv}.
The superluminal muon neutrinos with $\delta v_{\nu}$  in  Eq.~(\ref{eq:opr}) can
not have energy larger than about 5 GeV from pion decay process,
$\pi^+ \to \mu^+ \nu_{\mu}$ and $\mu \to \nu_{\mu} + e + {\bar \nu}_e$~\cite{Bi:2011nd}. 
Several solutions to these challenges have been proposed 
as well~\cite{AmelinoCamelia:2011bz, Li:2011rt, 
Evslin:2011vq, Lingli:2011kh, Mohanty:2011rm, Li:2011ad, Ling:2011re}. 

It is well known that the traditional Lorentz invariance can be superseded by two approaches.
In the first approach, there exists a preferred frame of reference for Lorentz symmetry 
breaking~\cite{Colladay:1996iz, Colladay:1996iz-B, Colladay:1996iz-C}. In
the second approach, the Lorentz invariance is 
deformed so that the principle of relativity of intertial frames is preserved,
while the Lorentz transformations, the energy-momentum relations, and the energy-momentum
conservation laws are deformed~\cite{AmelinoCamelia:2000mn, AmelinoCamelia:2000mn-B,
AmelinoCamelia:2000mn-C, AmelinoCamelia:2000mn-D, AmelinoCamelia:2000mn-E}. The above theoretical challenges
are valid in the first approach. Interestingly, the Bremsstrahlung 
processes may be forbidden and the pion decays may not be a problem if the Lorentz symmetry 
is deformed~\cite{AmelinoCamelia:2011bz}. 
 However, the studied model  
can not be consistent with the OPERA results obviously since $\delta v_{\nu}$ is 
proportional to the neutrino energy square~\cite{AmelinoCamelia:2011bz}.

In this paper, we study the constant $\delta v_{\nu}$ around
$10^{-6}$ in two scenarios.  To be consistent with the OPERA, Borexino, 
and ICARUS experiments and 
with the SN1987a observations~\cite{imb, imb-B, imb-C}, we assume that
$\delta v_{\nu}$ is about $ 10^{-6}$ on the Earth
and is smaller than about $2 \times 10^{-9}$ on the interstellar scale,
as suggested in the background dependent Lorentz violation
proposals~\cite{Dvali:2011mn, Alexandre:2011bu, Ciuffoli:2011ji, Li:2011zm}.
To explain the above theoretical challenges, 
we consider the deformed Lorentz invariance, and show that 
the superluminal neutrino dispersion relations can be realized 
properly while the modifications to the dispersion relations of 
the other Standard Model (SM) particles can be very tiny and negligible.
Moreover, we propose the
deformed energy and momentum conservation laws for a generic  
physical process.  In Scenario I the
momentum conservation law is preserved while the energy conservation law 
is deformed. In Scenario II the energy conservation law is preserved while
the momentum conservation law is deformed. 
We present the energy and momentum conservation laws in terms of
neutrino momentum in Scenario I and in terms of neutrino energy
in Scenario II. In such formats,  the energy and momentum conservation
laws are exactly the same as those in the traditional quantum 
field theory with Lorentz symmetry. Thus, all the above theoretical 
challenges can be automatically solved. To be concrete, we show that
 the Bremsstrahlung processes are forbidden and the pion decays
are not a problem.

{\bf {The Superluminal Neutrinos from Deformed Lorentz 
Invariance}~--}~Considering the effective field theory or string theory, we can parametrize 
the generic $\delta v_{\nu}$ for a neutrino as follows
\begin{eqnarray}
\delta v_{\nu} ~=~ -{{m_{\nu}^2}\over {2 P_{\nu}^2}} + \sum_{n\ge 0} a_n {{P_{\nu}^n}\over {M^n_*}}~,~\,
\label{DV-Generic}
\end{eqnarray}
where $m_{\nu}$ and $P_{\nu}\equiv |\vec{P}_{\nu}|$ are respectively the neutrino 
mass and momentum, $a_n$
are the coefficients, and $M_*$ is the effective normalization scale. Note that
the OPERA results have weak energy dependence, we can only consider the $a_0$ term
and $a_1$ term. The other terms must be very small if they are not vanish.
 The pure $a_1$ term can not be obtained in the 
Lorentz violation theory with CPT symmetry~\cite{Colladay:1996iz, Colladay:1996iz-B, Colladay:1996iz-C}. 
Interestingly, in the Type IIB 
string theory, we can obtain this term naturally by calculating the 
four-point function~\cite{Li:2011zm, Li:2009tt}. Especially, all the theoretical
challenges can be solved in such string scenario~\cite{Li:2011rt}. Thus,
in this Letter we will concentrate 
on the $a_0$ term, {\it i.e.}, the constant $\delta v_{\nu}$,
as suggested by the OPERA experiment.

With the deformed Lorentz invariance, we can parametrize the generic
 dispersion relation for the SM particles as follows
\begin{eqnarray}
E^2_{A} ~=~ \vec{P}_{A}^2 + m_{A}^2 + \xi^A_P \vec{P}_{A}^2 + \xi^A_{PE} P_{A} E_{A} +
\xi^A_E E^2_{A}~,~\,
\end{eqnarray}
where $E_{A}$, $\vec{P}_{A}$ and $m_A$ are respectively the energy,
momentum and mass for a SM particle $A$,
and $\xi^A_P$, $\xi^A_{PE}$, and $\xi^A_E$ are coefficients.
Because the SM particle masses are invariant under the deformed
Lorentz transformations, $\xi^A_P$, $\xi^A_{PE}$, and $\xi^A_E$ are universal
functions of $m_A$ for all the SM particles, {\it i.e.},
 $\xi^A_P \equiv \xi_P(m_A)$, $\xi^A_{PE} \equiv \xi_{PE}(m_A) $, 
and $\xi^A_E \equiv \xi_E(m_A)$.
From the effective field theory with CPT symmetry, 
we can obtain the $\xi^A_P \vec{P}_{A}^2$ or $\xi^A_E E^2_{A}$ term
independently~\cite{Giudice:2011mm, Colladay:1996iz, Colladay:1996iz-B, Colladay:1996iz-C}. 
If the $\xi^A_{PE} E_{A} P_{A}$ term exists, it must come from
the interference term. After redefining the kinetic terms
and mass terms for the SM particles,
we can have either the $\xi^A_P \vec{P}_{A}^2$ term or the $\xi^A_E E^2_{A}$ term. 
Thus, we shall study two scenarios: Scenario I with only the $\xi^A_P \vec{P}_{\nu}^2$ term
and Scenario II with only the $\xi^A_E E^2_{\nu}$ term. 

First, we consider the Scenario I where $\delta v_{\nu} = \xi^{\nu}_P/2$. 
For simplicity, we choose
\begin{eqnarray}
\xi^{A}_P ~=~ \alpha_P {{m_A^2 M_{\rm IR}^2}\over\displaystyle {m_A^4 + M_{\rm IR}^4}}~,~\,
\end{eqnarray}
where $\alpha_P$ is a coefficient, and $M_{\rm IR}$ is the infrared (IR) scale and
will be assumed to be the cosmological constant scale. 
Choosing $m_{\nu} = 0.05$~eV,
$\xi_P^{\nu}=1.0\times 10^{-6}$, and $M_{\rm IR}=2.3\times 10^{-3}$~eV, we get
$\alpha_P=4.73\times 10^{-4}$. Note that the electron mass is 0.511 MeV, we
have $\xi_P^e = 9.57 \times 10^{-21}$. It is easy to show that $\xi_P^A$ for 
all the rest SM particles are smaller than $\xi_P^e$. Thus, the modifications 
to the dispersion relations of the SM particles except the neutrinos are
very tiny and can be neglected.

The neutrino dispersion relation is required to be
invariant under the deformed boost generators $N^{\nu}_i$
\begin{eqnarray}
[N^{\nu}_i, ~ E^2_{\nu} - \vec{P}_{\nu}^2 - \xi^{\nu}_P \vec{P}_{\nu}^2 ]~=~0~.~\,
\end{eqnarray}
And the Einstein special relativity should be realized at the
$\xi^A_P=0$ limit. Thus, we obtain
\begin{eqnarray}
  [N^{\nu}_i, ~E_{\nu}] ~ = ~ \beta_1 (P_{\nu})_i~,~~
  [N^{\nu}_i, ~(P_{\nu})_j] ~ = ~ {1\over  {\beta_1}} E _{\nu}\delta_{ij}~,~\,
\end{eqnarray}
where $\beta_1 ~=~ {\sqrt {1+\xi^{\nu}_P}}$.

Let us consider a generic physical process: the initial states include
$n$ neutrinos and $n'$ other SM particles, and the final
states include $m$ neutrinos and $m'$ other SM particles.
We obtain the momentum and energy conservation laws  which
are invariant under the deformed Lorentz symmetry
\begin{eqnarray}
 \sum_{k=1}^n \vec{P}^i_{\nu k} + \sum_{k=1}^{n'} \vec{P}^i_{k} &=&
\sum_{k=1}^m \vec{P}^f_{\nu k} + \sum_{k=1}^{m'} \vec{P}^f_{k}~,~ 
\label{MOM-1}\\
 {1\over {\beta_1}} \sum_{k=1}^n E^i_{\nu k} + \sum_{k=1}^{n'} E^i_{k} &=&
   {1\over {\beta_1}} \sum_{k=1}^m E^f_{\nu k} + \sum_{k=1}^{m'} E^f_{k}~,~
\end{eqnarray}
where $P_k$ and $E_k$ are respectively the momentum and energy for
the other SM particles, and
the upper indices $i$ and $f$ denote the initial state
and final state, respectively.
Interestingly, the momentum conservation law is preserved as well.
Note that the neutrino masses are tiny and can be neglected, we get
$E_{\nu} = \beta_1 P_{\nu}$. Thus,
the energy conservation law can be rewritten as follows
\begin{eqnarray}
 \sum_{k=1}^n P^i_{\nu k} + \sum_{k=1}^{n'} E^i_{k} ~=~
   \sum_{k=1}^m P^f_{\nu k} + \sum_{k=1}^{m'} E^f_{k}~.~
\label{ENE-1}
\end{eqnarray}
Therefore, the momentum conservation law in Eq.~(\ref{MOM-1}) and
energy conservation law in Eq.~(\ref{ENE-1}) are the same
as those in the traditional quantume field theory with 
Lorentz symmetry. And then all the theoretical challenges can
be solved naturally.

Second, we consider the Scenario II where $\delta v_{\nu} = \xi^{\nu}_E/2$. 
Similar to the Scenario I, we choose
\begin{eqnarray}
\xi^{A}_E ~=~ \alpha_E {{m_A^2 M_{\rm IR}^2}\over\displaystyle {m_A^4 + M_{\rm IR}^4}}~,~\,
\end{eqnarray}
where  $\alpha_E =4.73\times 10^{-4}$. Thus, the modifications 
to the dispersion relations of the SM particles except the neutrinos are
very tiny and can be neglected.

The neutrino 
dispersion relation is required to be invariant under the deformed boost generators
\begin{eqnarray}
[N^{\nu}_i, ~ E^2_{\nu} - \vec{P}_{\nu}^2 - \xi^{\nu}_E E_{\nu}^2 ]~=~0~.~\,
\end{eqnarray}
Thus, we obtain
\begin{eqnarray}
  [N^{\nu}_i, ~E_{\nu}] ~ = ~   {1\over  {\beta_2}} (P_{\nu})_i~,~~
  [N^{\nu}_i, ~(P_{\nu})_j] ~ = ~  {\beta_2}   E_{\nu} \delta_{ij}~,~\,
\end{eqnarray}
where $\beta_2 ~=~ {\sqrt {1-\xi^{\nu}_E}}$.

For the generic physical process given in Scenario I,
we obtain the momentum and energy conservation laws which
are invariant under the deformed Lorentz symmetry
\begin{eqnarray}
 {1\over {\beta_2}} \sum_{k=1}^n \vec{P}^i_{\nu k} + \sum_{k=1}^{n'} \vec{P}^i_{k} &=&
{1\over {\beta_2}} \sum_{k=1}^m \vec{P}^f_{\nu k} + \sum_{k=1}^{m'} \vec{P}^f_{k}~,~ \\
 \sum_{k=1}^n E^i_{\nu k} + \sum_{k=1}^{n'} E^i_{k} &=&
    \sum_{k=1}^m E^f_{\nu k} + \sum_{k=1}^{m'} E^f_{k}~.~
\label{ENE-2}
\end{eqnarray}
Interestingly, the energy conservation law is preserved as well.
Note that the neutrino masses are tiny and can be neglected, we have
$P_{\nu} =  \beta_2 E_{\nu} $. Thus,
the momentum conservation law can be rewritten as follows
\begin{eqnarray}
\sum_{k=1}^n E^i_{\nu k} \vec{r}^{~i}_{\nu k} + \sum_{k=1}^{n'} \vec{P}^i_{k} &=&
 \sum_{k=1}^m E^f_{\nu k} \vec{r}^{~f}_{\nu k} + \sum_{k=1}^{m'} \vec{P}^f_{k}~,~
\label{MOM-2}
\end{eqnarray}
where 
\begin{eqnarray}
\vec{r}^{~i}_{\nu k} ~\equiv~{{\vec{P}^i_{\nu k}}\over {P^i_{\nu k}}}~,~~~
\vec{r}^{~f}_{\nu k} ~\equiv~{{\vec{P}^f_{\nu k}}\over {P^f_{\nu k}}}~.~
\end{eqnarray}
Therefore, the momentum conservation law in Eq.~(\ref{MOM-2}) and
energy conservation law in Eq.~(\ref{ENE-2}) are the same
as those in the traditional quantume field theory with 
Lorentz symmetry. And then all the theoretical challenges can
be solved naturally.

{\bf Theoretical Challenges~--}~We will consider the
theoretical challenges, and prove that the  Bremsstrahlung
processes~\cite{Cohen:2011hx} are forbidden and there is no
problem for pion decays~\cite{GonzalezMestres:2011jc, Bi:2011nd, Cowsik:2011wv}.

First, let us consider the Bremsstrahlung effects.
As an simple example, we show that the  most important process 
 $\nu_{\mu} \to \nu_{\mu} + e^+ + e^-$  is forbidden in 
Scenario I. From the generic 
energy-momentum conservation
laws in Eqs.~(\ref{MOM-1}) and (\ref{ENE-1}), we have
\begin{eqnarray}
  \vec{P}^i_{\nu}  &=&
 \vec{P}^f_{\nu k} +  \vec{P}^f_{e^+ } + \vec{P}^f_{e^- }~,~
\label{MOM-3} \\
 P^i_{\nu}  &=& P^f_{\nu } +  E^f_{e^+} +  E^f_{e^-}~.~
\label{ENE-3}
\end{eqnarray}

Let us suppose that this process is not forbidden.
As we know, the electron and positron masses are about 0.511 MeV,
 and the neutrino masses are around 0.05 eV.
From Eq.~(\ref{MOM-3}) we obtain
\begin{eqnarray}
(\vec{P}^i_{\nu}-\vec{P}^f_{\nu k})^2 ~=~ (\vec{P}^f_{e^+ } + \vec{P}^f_{e^- })^2
~<~ (E^f_{e^+ } + E^f_{e^- })^2~,~
\end{eqnarray}
where the above inequality is achieved by considering the electron and positron
masses. Using Eq.~(\ref{ENE-3}), we obtain 
\begin{eqnarray}
(\vec{P}^i_{\nu}-\vec{P}^f_{\nu k})^2 ~<~ ( P^i_{\nu}-P^f_{\nu })^2~.~
\end{eqnarray}
This inequality can not be satisfied obviously, thus, the 
process  $\nu_{\mu} \to \nu_{\mu} + e^+ + e^-$ is indeed forbidden.

Second, let us consider the pion decays via the process  
$\pi^+ \to \mu^+ + \nu_{\mu}$. Similar to the 
Refs.~\cite{GonzalezMestres:2011jc, Bi:2011nd, Cowsik:2011wv},
we consider the neutrino dispersion relation in Scenario I.
In Refs.~\cite{GonzalezMestres:2011jc, Bi:2011nd}, 
using the preferred frame of reference, the authors obtained
their results by assuming the following threshold condition
\begin{eqnarray}
m_{\pi}  \ge m_{\mu} + \sqrt{\xi^{\nu}_P} P_{\nu}~.~
\end{eqnarray}
However, this threshold condition is not valid 
in the deformed Lorentz invariance.

In addition, in Ref.~\cite{Cowsik:2011wv}, the authors obtained their results
 by assuming the following energy and momentum conservation laws
\begin{eqnarray}
\vec{P}_{\pi} ~=~ \vec{P}_{\mu} + \vec{P}_{\nu}~,~~~E_{\pi} ~=~ E_{\mu} + 
{\sqrt {1+\xi^{\nu}_P}} P_{\nu}~.~
\end{eqnarray}
However, such laws are obviously different from the generic 
energy and momentum conservation laws
in Eqs.~(\ref{MOM-1}) and (\ref{ENE-1}) for
the deformed Lorentz invariance. In particular,
the conservation laws in Eqs.~(\ref{MOM-1}) and (\ref{ENE-1}) have no
$\xi^{\nu}_P$ dependence, thus, we do not have the severe difficulties
with the kinematics of the pion decays.


{\bf Conclusion~--}~We studied two superluminal neutrino scenarios where $\delta v$ 
is a constant.  To be consistent with the OPERA, Borexino, and ICARUS experiments 
and with the SN1987a observations, we assumed that $\delta v_{\nu}$ 
on the Earth is much larger than that on the interstellar scale.
To explain the theoretical challenges, we considered the 
deformed Lorentz invariance, and showed that 
the superluminal neutrino dispersion relations can be realized 
properly while the modifications to the dispersion relations of 
the other SM particles can be negligible.
In addition, we
proposed the deformed energy and momentum conservation laws for 
a generic physical process. In Scenario I the momentum conservation 
law is preserved while the energy conservation law is deformed. 
In Scenario II the energy conservation law is preserved while
the momentum conservation law is deformed. We presented the energy and 
momentum conservation laws in terms of neutrino momentum in 
Scenario I and in terms of neutrino energy in Scenario II. 
In such formats, the energy and momentum conservation
laws are exactly the same as those in the traditional quantum 
field theory with Lorentz symmetry. Thus, all the theoretical 
challenges can be automatically solved. To be concrete, we showed that
the Bremsstrahlung processes are forbidden and the pion decays
are not a problem.

{\bf Acknowledgments~--}~This research was supported in part 
by the Natural Science Foundation of China 
under grant numbers 10821504 and 11075194 (YH, TL and YQ), 
10975078 and 11025525 (YL), and 11105124 (FW),
and by the DOE grant DE-FG03-95-Er-40917 (TL and DVN).



\begin{thebibliography}{99}













\bibitem{Adam:2011zb} 
  T.~Adam {\it et al.}  [OPERA Collaboration],
  arXiv:1109.4897 [hep-ex].


\bibitem{AlvarezSanchez:2012wg} 
  P.~Alvarez Sanchez {\it et al.}  [Borexino Collaboration],
  arXiv:1207.6860 [hep-ex].

\bibitem{Antonello:2012be} 
  M.~Antonello, B.~Baibussinov, P.~Benetti, E.~Calligarich, N.~Canci, S.~Centro, A.~Cesana and K.~Cieslik {\it et al.},
  arXiv:1208.2629 [hep-ex].

\bibitem{Ellis:2008fc} 
  J.~R.~Ellis, N.~Harries, A.~Meregaglia, A.~Rubbia and A.~Sakharov,
  Phys.\ Rev.\ D {\bf 78}, 033013 (2008)
  [arXiv:0805.0253 [hep-ph]].




\bibitem{Cacciapaglia:2011ax} 
  G.~Cacciapaglia, A.~Deandrea and L.~Panizzi,
  JHEP {\bf 1111}, 137 (2011)
  [arXiv:1109.4980 [hep-ph]].


\bibitem{AmelinoCamelia:2011dx} 
  G.~Amelino-Camelia, G.~Gubitosi, N.~Loret, F.~Mercati, G.~Rosati and P.~Lipari,
  Int.\ J.\ Mod.\ Phys.\ D {\bf 20}, 2623 (2011)
  [arXiv:1109.5172 [hep-ph]].

\bibitem{Giudice:2011mm} 
  G.~F.~Giudice, S.~Sibiryakov and A.~Strumia,
  Nucl.\ Phys.\ B {\bf 861}, 1 (2012)
  [arXiv:1109.5682 [hep-ph]].


\bibitem{Dvali:2011mn} 
  G.~Dvali and A.~Vikman,
  JHEP {\bf 1202}, 134 (2012)
  [arXiv:1109.5685 [hep-ph]].



\bibitem{Mann:2011rd}
  R.~B.~Mann, U.~Sarkar,
  arXiv:1109.5749 [hep-ph].


\bibitem{Drago:2011ua} 
  A.~Drago, I.~Masina, G.~Pagliara and R.~Tripiccione,
  Europhys.\ Lett.\  {\bf 97}, 21002 (2012)
  [arXiv:1109.5917 [hep-ph]].


\bibitem{Li:2011ue}
  M.~Li, T.~Wang,
  arXiv:1109.5924 [hep-ph].

\bibitem{Pfeifer:2011ve}
  C.~Pfeifer, M.~N.~R.~Wohlfarth,
  arXiv:1109.6005 [gr-qc].


\bibitem{Lingli:2011yn}
  Z.~Lingli, B.~-Q.~Ma,
  arXiv:1109.6097 [hep-ph].


\bibitem{Alexandre:2011bu} 
  J.~Alexandre, J.~Ellis and N.~E.~Mavromatos,
  Phys.\ Lett.\ B {\bf 706}, 456 (2012)
  [arXiv:1109.6296 [hep-ph]].


\bibitem{Cohen:2011hx} 
  A.~G.~Cohen and S.~L.~Glashow,
  Phys.\ Rev.\ Lett.\  {\bf 107}, 181803 (2011)
  [arXiv:1109.6562 [hep-ph]].


\bibitem{GonzalezMestres:2011jc}
  L.~Gonzalez-Mestres,
  arXiv:1109.6630 [physics.gen-ph].


\bibitem{Matone:2011jd}
  M.~Matone,
  arXiv:1109.6631 [hep-ph].

\bibitem{Ciuffoli:2011ji}
  E.~Ciuffoli, J.~Evslin, J.~Liu, X.~Zhang,
  arXiv:1109.6641 [hep-ph].


\bibitem{Bi:2011nd} 
  X.~-J.~Bi, P.~-F.~Yin, Z.~-H.~Yu and Q.~Yuan,
  Phys.\ Rev.\ Lett.\  {\bf 107}, 241802 (2011)
  [arXiv:1109.6667 [hep-ph]].


\bibitem{Wang:2011sz}
  P.~Wang, H.~Wu, H.~Yang,
  arXiv:1109.6930 [hep-ph].

\bibitem{Wang:2011sz-B}
P.~Wang, H.~Wu, H.~Yang,
  arXiv:1110.0449 [hep-ph].


\bibitem{Cowsik:2011wv} 
  R.~Cowsik, S.~Nussinov and U.~Sarkar,
  Phys.\ Rev.\ Lett.\  {\bf 107}, 251801 (2011)
  [arXiv:1110.0241 [hep-ph]].


\bibitem{Li:2011zm}
  T.~Li, D.~V.~Nanopoulos,
  arXiv:1110.0451 [hep-ph].



\bibitem{AmelinoCamelia:2011bz} 
  G.~Amelino-Camelia, L.~Freidel, J.~Kowalski-Glikman and L.~Smolin,
  Mod.\ Phys.\ Lett.\ A {\bf 27}, 1250063 (2012)
  [arXiv:1110.0521 [hep-ph]].


\bibitem{Moffat:2011ue}
  J.~W.~Moffat,
  arXiv:1110.1330 [hep-ph].

\bibitem{Faraggi:2011en}
  A.~E.~Faraggi,
  arXiv:1110.1857 [hep-ph].

\bibitem{Konoplya:2011hf} 
  R.~A.~Konoplya and A.~Zhidenko,
  Phys.\ Rev.\ D {\bf 86}, 023531 (2012)
  [arXiv:1110.2015 [hep-th]].

\bibitem{Li:2011rt} 
  T.~Li and D.~V.~Nanopoulos,
  Eur.\ Phys.\ J.\ C {\bf 72}, 2044 (2012)
  [arXiv:1110.3451 [hep-ph]].


\bibitem{Zhao:2011sb} 
  L.~-A.~Zhao and X.~Zhang,
  arXiv:1110.6577 [hep-ph].


\bibitem{Chang:2011td} 
  Z.~Chang, X.~Li and S.~Wang,
  Mod.\ Phys.\ Lett.\  {\bf 27}, 1250058 (2012)
  [arXiv:1110.6673 [hep-ph]].


\bibitem{Almeida:2011ca} 
  C.~A.~G.~Almeida, M.~A.~Anacleto, F.~A.~Brito and E.~Passos,
  Eur.\ Phys.\ J.\ C {\bf 72}, 1855 (2012)
  [arXiv:1111.0093 [hep-ph]].

\bibitem{Matone:2011fn} 
  M.~Matone,
  arXiv:1111.0270 [hep-ph].

\bibitem{Alles:2011wq} 
  B.~Alles,
  Phys.\ Rev.\ D {\bf 85}, 047501 (2012)
  [arXiv:1111.0805 [hep-ph]].

\bibitem{Evslin:2011vq} 
  J.~Evslin,
  arXiv:1111.0733 [hep-ph].

\bibitem{Lingli:2011kh} 
  Z.~Lingli and B.~-Q.~Ma,
  arXiv:1111.1574 [hep-ph].

\bibitem{Mohanty:2011rm} 
  S.~Mohanty and S.~Rao,
  arXiv:1111.2725 [hep-ph].

\bibitem{Li:2011ad} 
  M.~Li, D.~Liu, J.~Meng, T.~Wang and L.~Zhou,
  arXiv:1111.3294 [hep-ph].

\bibitem{Ling:2011re} 
  Y.~Ling,
  arXiv:1111.3716 [hep-ph].





\bibitem{ICARUS:2011aa} 
  M.~Antonello {\it et al.}  [ICARUS Collaboration],
  Phys.\ Lett.\ B {\bf 711}, 270 (2012)
  [arXiv:1110.3763 [hep-ex]].



\bibitem{Colladay:1996iz}
  D.~Colladay, V.~A.~Kostelecky,
  Phys.\ Rev.\  {\bf D55}, 6760-6774 (1997).

\bibitem{Colladay:1996iz-B}
 G.~Amelino-Camelia, J.~R.~Ellis, N.~E.~Mavromatos, D.~V.~Nanopoulos, S.~Sarkar,
  Nature {\bf 393}, 763-765 (1998).

\bibitem{Colladay:1996iz-C}
S.~R.~Coleman, S.~L.~Glashow,
  Phys.\ Rev.\  {\bf D59}, 116008 (1999).


\bibitem{AmelinoCamelia:2000mn}
  G.~Amelino-Camelia,
  Int.\ J.\ Mod.\ Phys.\  {\bf D11}, 35-60 (2002).


\bibitem{AmelinoCamelia:2000mn-B}
  G.~Amelino-Camelia,
  Phys.\ Lett.\  {\bf B510}, 255-263 (2001).

\bibitem{AmelinoCamelia:2000mn-C}
J.~Kowalski-Glikman,
  Phys.\ Lett.\  {\bf A286}, 391-394 (2001).

\bibitem{AmelinoCamelia:2000mn-D}
J.~Magueijo, L.~Smolin,
  Phys.\ Rev.\ Lett.\  {\bf 88}, 190403 (2002).

\bibitem{AmelinoCamelia:2000mn-E}
  Phys.\ Rev.\  {\bf D67}, 044017 (2003).


\bibitem{imb}
  R.~M.~Bionta {\it et al.} [IMB Collaboration],
  Phys.\ Rev.\ Lett.\  {58} (1987) 1494.

\bibitem{imb-B}
E.~N.~Alekseev, L.~N.~Alekseeva, I.~V.~Krivosheina and V.~I.~Volchenko,
  Phys.\ Lett.\  B {205} (1988) 209.

\bibitem{imb-C}
 K.~Hirata {\it et al.}  [KAMIOKANDE-II Collaboration],
  Phys.\ Rev.\ Lett.\  {58} (1987) 1490.



\bibitem{Li:2009tt}
  T.~Li, N.~E.~Mavromatos, D.~V.~Nanopoulos, D.~Xie,
  Phys.\ Lett.\  {\bf B679}, 407-413 (2009).
































\end{thebibliography}
\end{document}